\documentstyle[12pt,aasms]{article} 

\def\chaphead{}

\def\spose#1{\hbox to 0pt{#1\hss}}
     
\def\={\overline}

\newcount\notenumber
\newcount\eqnumber
\newcount\fignumber
\newbox\abstr
\newbox\figca

\def\etal{{\it et al. }}
\def\eg{{\it e.g., }}
\def\ie{{\it i.e., }}

\def\note#1{\footnote{$^{\the\notenumber}$}{#1}\global\advance\notenumber by 1}
\def\foot#1{\raise3pt\hbox{\eightrm \the\notenumber}
     \hfil\par\vskip3pt\hrule\vskip6pt
     \noindent\raise3pt\hbox{\eightrm \the\notenumber}
     #1\par\vskip6pt\hrule\vskip3pt\noindent\global\advance\notenumber by 1}

\def\Dt{\spose{\raise 1.5ex\hbox{\hskip3pt$\mathchar"201$}}}    
\def\dt{\spose{\raise 1.0ex\hbox{\hskip2pt$\mathchar"201$}}}    

\def\new{{\rm\chaphead\the\eqnumber}\global\advance\eqnumber by 1}
\def\ref#1{\advance\eqnumber by -#1 \chaphead\the\eqnumber
     \advance\eqnumber by #1 }
\def\last{\advance\eqnumber by -1 {\rm\chaphead\the\eqnumber}\advance
     \eqnumber by 1}
\def\eqnam#1{\xdef#1{\chaphead\the\eqnumber}}
     
\def\nfig{\chaphead\the\fignumber\global\advance\fignumber by 1}
\def\nfiga#1{\chaphead\the\fignumber{#1}\global\advance\fignumber by 1}
\def\rfig#1{\advance\fignumber by -#1 \chaphead\the\fignumber
     \advance\fignumber by #1}
\def\refindent{\par\noindent\parskip=4pt\hangindent=3pc\hangafter=1 }

\def\apj#1#2#3{\refindent#1,  {ApJ,\ }{#2}, #3}

\def\mn#1#2#3{\refindent#1,  { MNRAS,\ }{#2}, #3}

\def\annrev#1#2#3{\refindent#1, { ARA \& A,\ }
{\bf2}, #3}

\def\refbook#1{\refindent#1}

\def\ltsim{\mathrel{\spose{\lower 3pt\hbox{$\mathchar"218$}}
     \raise 2.0pt\hbox{$\mathchar"13C$}}}
\def\gtsim{\mathrel{\spose{\lower 3pt\hbox{$\mathchar"218$}}
     \raise 2.0pt\hbox{$\mathchar"13E$}}}
     
\def\apequal{\mathrel{\spose{\lower 1pt\hbox{$\mathchar"218$}}
     \raise 2.0pt\hbox{$\mathchar"218$}}}
\newbox\grsign \setbox\grsign=\hbox{$>$} \newdimen\grdimen \grdimen=\ht\grsign
\newbox\simlessbox \newbox\simgreatbox
\setbox\simgreatbox=\hbox{\raise.5ex\hbox{$>$}\llap
     {\lower.5ex\hbox{$\sim$}}}\ht1=\grdimen\dp1=0pt
\setbox\simlessbox=\hbox{\raise.5ex\hbox{$<$}\llap
     {\lower.5ex\hbox{$\sim$}}}\ht2=\grdimen\dp2=0pt
\def\gtorder{\mathrel{\copy\simgreatbox}}
\def\ltorder{\mathrel{\copy\simlessbox}}

\def\eg{{\it e.g.,\ }}
\def\ie{{\it i.e.,\ }}
\def\etal{{\it et al.\ }}
\begin{document} 
\title{\bf Anisotropies in the Distribution of Satellite Galaxies}

\author{Dennis Zaritsky}
\affil{UCO/Lick Observatories and Board of Astronomy and
Astrophysics,} 
\affil{Univ. of California at Santa Cruz, Santa Cruz, CA, 95064}
\author{Rodney Smith}
\affil{Department of Physics, P.O. Box 913, University of Wales, 
College of Cardiff,}
\affil{Cardiff CF2 3YB, Wales.}
\author{Carlos S. Frenk} 
\affil{Department of Physics, South Road, University of Durham, Durham,
DH1 3LE, England}
\author{and} 
\author{Simon D.M. White}
\affil{Max-Planck-Institut f\"ur Astrophysik, Karl-Schwarzschild-Strasse
1,}
\affil{D-85748 Garching bei M\"unchen, Germany}

\abstract{
We find that satellites of isolated disk galaxies at projected
radii between 300 and 500 kpc are distributed
asymmetrically about the parent galaxy and aligned
preferentially with the disk minor axis.  The dynamical timescale at
these radii is sufficiently long that the shape of this distribution 
must reflect the formation history of the outer halo rather than its
internal evolution. We also find that the orbital angular momenta of
satellites at projected major axis distances of $\ltsim$ 200 kpc
tend to align with that of the central disk. These results demonstrate
that satellites are dynamically related to their primary galaxy. 
Satellites drawn from current simulations of hierarchical galaxy
formation exhibit neither the systematic alignment nor
the net rotation with the central disk that we find in the data. }

\section{Introduction}

Although it has become increasingly evident over the last two decades
that isolated
galaxies are surrounded by large and massive unseen halos (Faber 
\& Gallagher 1979; Zaritsky \& White 1994), the mass distribution 
in these halos remains weakly constrained.
Dark matter is generally thought to 
constitute between 90 and 99.5\% of the mass in the universe
and a thorough understanding of its distribution
is critical to the study of galaxy evolution and cosmology. 
The assumption that the halos of galaxies are spherical at large 
radii is often made, but has little observational or
theoretical support.  Satellite galaxies provide the best available test
of this assumption.

Null or marginal detections of anisotropy in
the relative orientation of galaxies (Hawley \& Peebles 1975; Sharp
\etal 1979; MacGillivray \etal 1982) have mostly suggested 
that galaxies are distributed
uniformly in angle about each other. A notable exception is the
study of Holmberg (1969) which found that
satellite galaxies at projected radii $< 50$ kpc tend to congregate
near the poles of their spiral primaries. 
Because this anisotropy was seen at small projected radii,
it has most often been interpreted as the result of 
selection effects or dynamical evolution. However, neither
of these alternatives appears able to account for
the magnitude of the effect (Quinn \& Goodman 1986). 
In this {\it Letter}, we present evidence for an anisotropic
distribution of satellite galaxies that is similar to that seen by Holmberg,
but which exists at much
larger radii, where neither selection biases nor dynamical evolution
seem a plausible explanation.

\vfill\eject
\section{Data and Analysis}

\subsection{The Spatial Distribution of Satellite Galaxies}

Our data consist of (1) projected positions and radial velocities for
115 satellite galaxies of 69 nearby (distance $<$ 100 Mpc,
for H$_0 =$ 75 km s$^{-1}$ Mpc$^{-1}$ as assumed throughout) isolated 
disk galaxies (the primaries) similar in luminosity to our own Galaxy,
and (2) a measurement
of the rotation sense of the disk for 57 of these primaries.
The selection of 
primary galaxies, the identification of satellites, and
all complementary observations are
discussed elsewhere (Zaritsky \etal 1993; Zaritsky \etal 1996). 
Satellites are defined to have
projected separations from their primaries that
are less than 500 kpc and velocities relative to their
primaries that are less than 500 km s$^{-1}$. In Figure 1, we show
the satellite distribution projected on the sky and oriented so that 
each primary is located at the origin with its
major axis placed along the horizontal axis.

The elongation of
the satellite distribution along the disk minor axis is evident 
both in the full dataset (left panel) and for
the 72 satellites of the 48 primaries that are
inclined at least 45$^\circ$ to the line-of-sight (right panel).
Observational biases due either to the telescope or to the instruments
cannot be responsible for this anisotropy because the position angles
of the primaries are randomly distributed on the sky and were not
taken into account when setting up the observations. Furthermore, our
data come from a variety of telescopes and instruments, as well as from the 
literature. The apparent anisotropy is qualitatively similar to that
found by
Holmberg (1969), but is occurring at projected 
radii much larger than the 50 kpc
limit of his dataset. We find no evidence for a polar
alignment of satellites 
at projected radii $<$ 200 kpc, and the visual impression
is that the satellite distribution is flattened in the opposite
sense along the disk major 
axis at these radii. This apparent flattening is not statistically 
significant. Finally, we cannot test Holmberg's result because we
have only 9 satellites at $r_p < 50$ kpc

We estimate the statistical significance of the polar alignment at large
projected radii by applying a 
Kolmogorov-Smirnoff (K-S) statistic to compare the distribution of
position angles relative to the disk major axis 
(reduced to the range 0 to 90$^\circ$) with the uniform
distribution expected for circular symmetry. We begin with the twenty 
satellites at the largest projected separations. We apply the K-S
test. We then add satellites one at a time,
stepping radially inward, and reapplying the K-S test to each enlarged
subsample. The results as a function of the lower limit in projected 
radius are shown in Figure 2 both for the full sample of 115 satellites
(solid line) and for a 
restricted sample of 72 satellites, for which the primary is inclined
by at least 45$^\circ$ to the line-of-sight (dashed line).
In both cases, the anisotropy is significant with greater than 99\% 
confidence for projected separations $\gtorder$ 250 kpc,  
and is absent for $r_p \ltorder 200$ kpc. Note that about a third of
our sample lies beyond 250 kpc, and about half lies within 200 kpc.
The apparent axial ratio of the satellite distribution in the outer regions
is roughly 5:3. A polar alignment of satellites is also observed
among the satellites of our own Galaxy (Hartwick 1996), although at a much
lower significance level.

These K-S confidence levels assume that each satellite is
independently drawn from the parent distribution. We now test whether
our conclusions are affected by 
spatial correlations among satellites.
Such correlations could arise because interlopers
(\ie galaxies projected along the line of sight, but not physically
associated with the primary) are clustered. We proceed
by randomly rejecting all but one satellite at $r_p > 250$ kpc for each 
primary-satellite system and then repeating the analysis.
The results are shown in Figure 2 as a dotted line and are fully
consistent with those found for the original sample. We conclude that
the anisotropy is real.

\subsection{The Net Rotation of the Satellite Population}

Our data enable us to determine whether 
the satellite ensemble rotates with the central galaxy disk.
Using the disk major axis and rotation direction to align the 
57 systems for which rotation has been measured,
we estimate the ensemble mean rotation by
averaging the line-of-sight velocity difference between satellite and
primary using opposite signs on the two sides of the projected
rotation axis.  We find a net rotation velocity of $34 \pm 14$ km
s$^{-1}$ in the
same sense as the disk; the data are shown in Figure 3.
The true mean rotation velocity of the satellite ensemble is
presumably larger
than this because the primary disks are observed at a range of
inclinations. Correcting, using the mean inclination of the primary disks
(54$^\circ$), gives an ``edge-on'' mean rotation velocity of 
 $v_{ROT} =$ 42 km s$^{-1}$. 

Although the satellite population rotates systematically, the
dispersion about the mean motion is large. Adopting a flat ``rotation
curve'' we measure a dispersion of 136 km s$^{-1}$,
implying $v_{ROT}/\sigma = 0.31$. This ratio is similar both to that
observed in giant elliptical galaxies (cf. van der Marel 1991), and to that 
measured for dark halos in N-body simulations (Frenk \etal 1988, 
Cole \& Lacey 1996). It is, however, quite unusual, either for
galaxies or for
simulated halos, to find such values when rotation is about
the minor axis. In N-body
simulations there is generally a poor correlation between the angular
momenta of the inner and outer regions of a halo. 

\section{Discussion}

Three possibilities suggest themselves for connecting
the observed satellites to the underlying
dark matter: (1) the satellites trace the dark matter,
that is their spatial and kinematic distributions can be considered a 
Poisson sampling from the corresponding dark matter distributions; (2)
the satellite and dark matter 
distributions differ because the satellite population has evolved
dynamically from its initial state (\eg dynamical friction has 
preferentially removed satellites on prograde or
plunging orbits); (3) the satellites do not trace
the dark matter because they formed in a biased way relative
to the dark matter.
There are at least two arguments against the second option.
The orbital timescale for distant satellites is a large fraction
of the age of the universe so that such strong dynamical
separation seems
unlikely. In addition, processes like tidal disruption and dynamical
friction should be stronger at small radii where the
dynamical times are shorter, whereas we see anisotropy only at
$r_p > 250$ kpc. The third option, biased galaxy formation,
may indeed lead to differences between the distributions of satellites 
and of dark matter, but current models give no reason to suspect
that the elongation of the satellite system should be perpendicular
to that of the dark matter. Bias appears to enhance the
contrast of the filaments and sheets already present in the dark matter 
distribution (\eg White \etal 1987). In effect, we are therefore left
with the first option, that the elongation of the outer satellite
distribution reflects that of the mass. The difficulty here is to
understand why the outer halo should be systematically elongated
{\it perpendicular} to the central disk.
 
The anisotropy we have detected would seem most consistent
with a near prolate distribution of satellites at large radii.
The apparent axial ratio of the satellite distribution is quite
consistent with the typical values found for dark matter halos in 
cosmological simulations, where 1:0.8:0.65 are representative values 
near the halo virial radius (corresponding to $\sim 180$ kpc for our 
galaxies) and more extreme values are common at larger radii 
(Efstathiou \etal 1988; Frenk \etal 1988;
Cole \& Lacey 1996). Note, however, that most
studies have found that in the standard hierarchical 
model the inner and outer regions of halos are only weakly
related; furthermore, rotation, when significant, is almost always
around one of the shortest axes  (\eg Barnes \& Efstathiou 1987; Frenk
\etal 1988; Warren \etal 1992). The rotation sense of our satellite
population and the orientation of the central galaxy are thus both
puzzling in the context of these models.

To explore this further, we constructed a sample of simulated
satellite galaxies drawn from the simulations of
Navarro, Frenk, \& White (1994, 1995). The sample was chosen 
from simulations of the evolution of isolated disk galaxies using
the selection criteria of our observational dataset. The
thirty simulations, each typically containing 1 to 2 satellites, are
viewed from random angles in order to produce pseudo-observational samples of
95 satellites (corresponding to our sample of real satellites of primaries
with known rotation sense). None 
of the five samples we constructed showed any sign of
a significant deviation from circular symmetry when the ``observations''
were superposed and aligned in the same manner as the real data.
For these simulations, at least, the correlation between galaxy
angular momentum and satellite position is too weak to reproduce our 
observational result. It is important to note that, as expected, the 
outer halos in these simulations have elongated shapes which align with
nearby structures. The problem is that the orientation of the central
disk appears uncorrelated with these preferred directions.

We also examine whether the simulations produce a coherent rotation of
the satellite ensemble when they are superposed in the same way as
the real data. We calculate the
correlation between major axis projected distance and
satellite-primary velocity difference for our five artificial
samples, but in no case do we find a significant correlation or an
apparent rotation exceeding 10 km s$^{-1}$. In contrast, the
correlation for the real 
data, shown in Figure 3, is significant at the 99.4\% level (Spearman
rank coefficient = 0.349) while the net rotation is 34 km s$^{-1}$.
Once again the angular momentum of the central galaxies in the
simulations appears to be more weakly related to the properties of the
satellite population than is the case in the real data.  

Navarro \etal (1995) already
noted that angular momentum losses during formation of these
simulated disks were too great for the disks to be compatible with
observed spirals. They suggested a possible resolution of this
problem -- that feedback from early star formation might keep much of
the gas hot, allowing it to settle later and with less angular
momentum loss into the disk. If this indeed happened, the properties
of the outer halo might plausibly be more faithfully reflected in the
disk than they are in the simulations. The correlation
between the angular momentum of the disk and that of the satellite population
might then be understood. The apparent elongation along the rotation
axis would, however, remain a puzzle. 

In summary, we find that the satellite distribution in the outer
halo regions of isolated spiral galaxies is elongated in the direction
perpendicular to the stellar disk. In addition the satellite
population tends to rotate in the same sense as the central disk
even though its typical physical scale is much larger (200 kpc as
opposed to 5 kpc). The rotation corresponds to $v/\sigma\sim 0.3$ for
the satellite system when the central disks are seen edge-on.
Current simulations of galaxy formation do not reproduce these
regularities. There are still some
missing pieces in our understanding of the galaxy formation process.

\vskip 1in
\noindent
ACKNOWLEDGMENTS: DZ acknowledges partial financial support from NASA through
HF-1027.01-91A, a grant from the California Space Institute, and HST
grant AR-06370.01-95A. CSF acknowledges receipt of a PPARC Senior
Research Fellowship. 
\vfill\eject
\vskip 1cm
\noindent
{\centerline{\bf References}}

\bigskip
\apj{Barnes, J., \& Efstathiou, G. 1987}{319}{575}

\mn{Cole, S., \& Lacey, C. 1996}{281}{716}

\mn{Efstathiou, G., Frenk, C.S., White, S.D.M., \& Davis, M. 1988}
{235}{715}

\apj{Frenk, C.S., White, S.D.M., Efstathiou, G. \& Davis, M. 1988}{327}{507}

\annrev{Faber, S.M., \& Gallagher, J.S. 1979}
{17}, {135}

\refbook{Hartwick, F.D.A. 1996 in {\it 
Formation of the Galactic Halo... Inside and Out}, ASP Conf. Series
vol 92, eds. H. Morrison \& A. Sarajedini, 444}

\refbook{Holmberg, E. 1969, {Ark. Astron.,} {\bf 5}, 305}

\mn{Navarro, J.F., Frenk, C.S., \& White, S.D.M. 1994}{267}{L1}

\mn{Navarro, J.F., Frenk, C.S., \& White, S.D.M. 1995}{275}{56}

\apj{Quinn, P.J. \& Goodman, J. 1986}{309}{472}

\mn{Sharp, N.A., Lin, D.N.C., \& White, S.D.M. 1979}{187}{287}

\mn{van den Marel, R.P. 1991}{253}{710}

\apj{Warren, M.S., Quinn, P.J., Salmon, J.K., \& Zurek, W.H. 1992}{399}{405}

\apj{White, S.D.M., Frenk, C.S., Davis, M., \& Efstathiou, G. 1987}{313}{505}

\apj{Zaritsky, D., Smith, R., Frenk, C., \& White, S.D.M. 1993}
{405}{464}

\refbook{Zaritsky, D., Smith, R., Frenk, C., \& White, S.D.M. 1996,
ApJ, in press}.

\apj{Zaritsky, D., \& White, S.D.M. 1994}{435}{599}.

\vfill\eject
{\centerline{\bf Figure Captions}}

\bigskip
\noindent
1) The projected distribution of satellite galaxies referenced to a 
primary galaxy centered on the origin with its major axis aligned with
the horizontal. The left panel includes all 115 satellites.
The right panel includes the 72 satellites of the 48 primaries that
are inclined more than 45$^{\circ}$ with respect to the line-of-
sight.

\bigskip
\noindent
2) The results of a K-S test comparison between the observed and
a uniform azimuthal distribution of satellites. The sample
lower boundary is set in projected separation and is plotted along
the x axis. Probability
refers to the probability that the observed azimuthal distribution
is not drawn from
a uniform distribution. The solid line 
represents the full sample of 115 satellites, 
the dashed line represents the sample of 72 satellites for which
the the primary disk is inclined at least 45$^\circ$ to the
line-of-sight, and the dotted line represents the sample
where multiple satellite groups have been removed (see text).

\bigskip
\noindent
3) The rotation curve of the satellite system along the disk
major axis. The lower panel includes the data 
for the 61 satellites of primaries inclined more than 45$^{\circ}$ 
to the line-of-sight for which we have a disk rotation measurement. 
Each system is oriented so that
the primary disk at positive radius has 
positive velocity. The upper panel shows a running mean (dashed
line) and median (solid line) of the nearest
nine satellite galaxy velocities.

\begin{figure}
\caption{}
\plotone{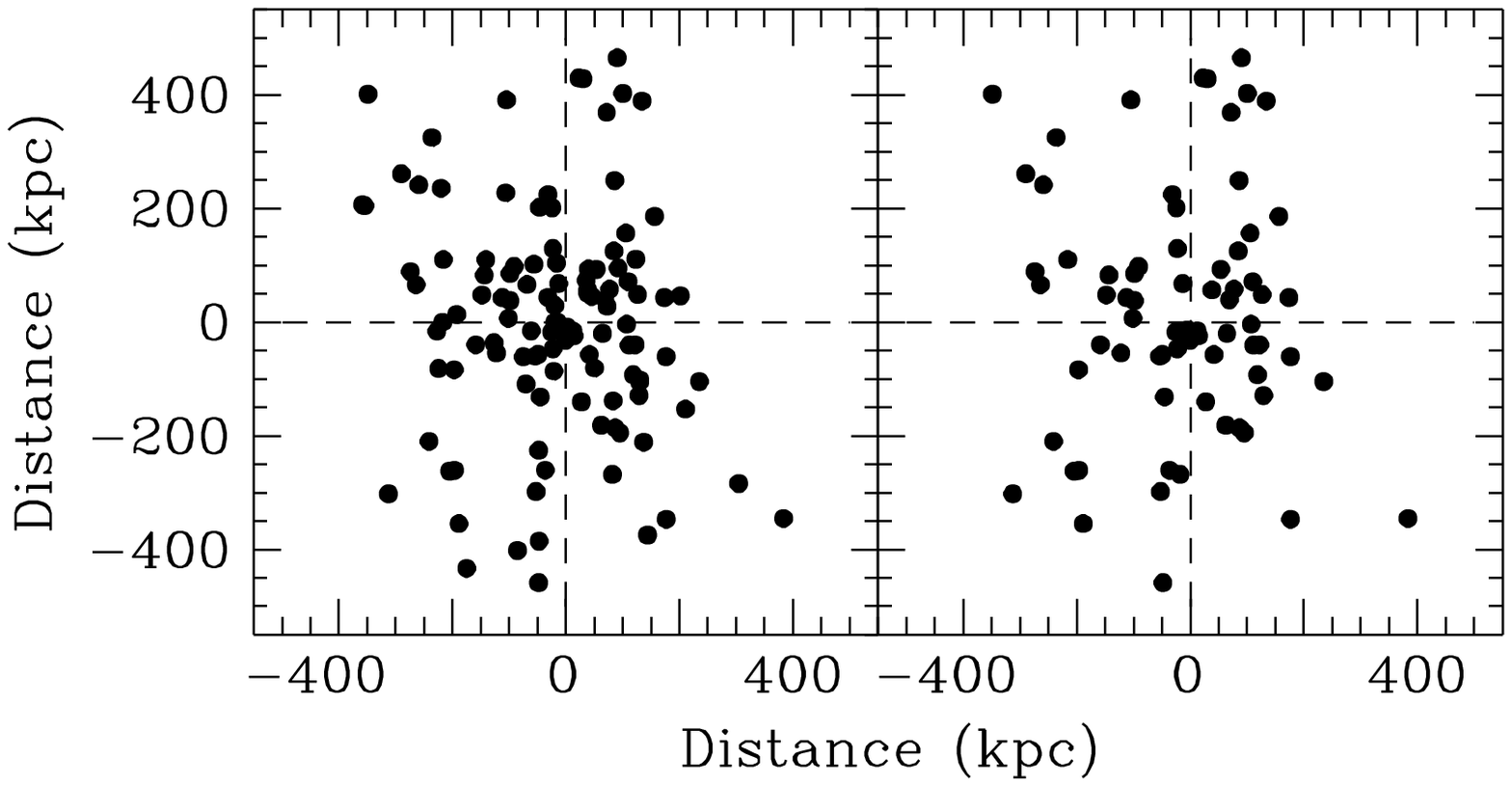}
\end{figure}

\begin{figure}
\caption{}
\plotone{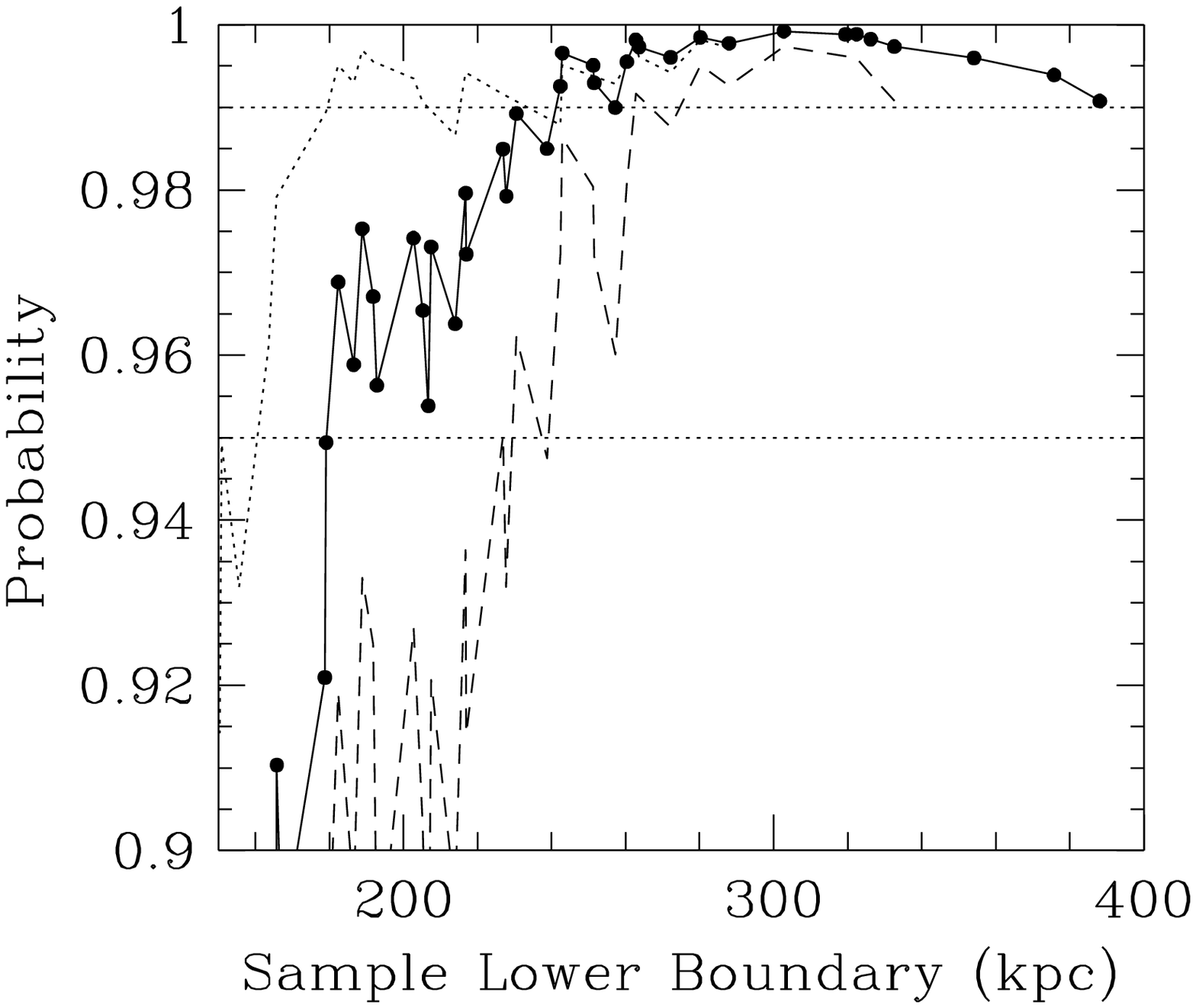}
\end{figure}

\begin{figure}
\caption{}
\plotone{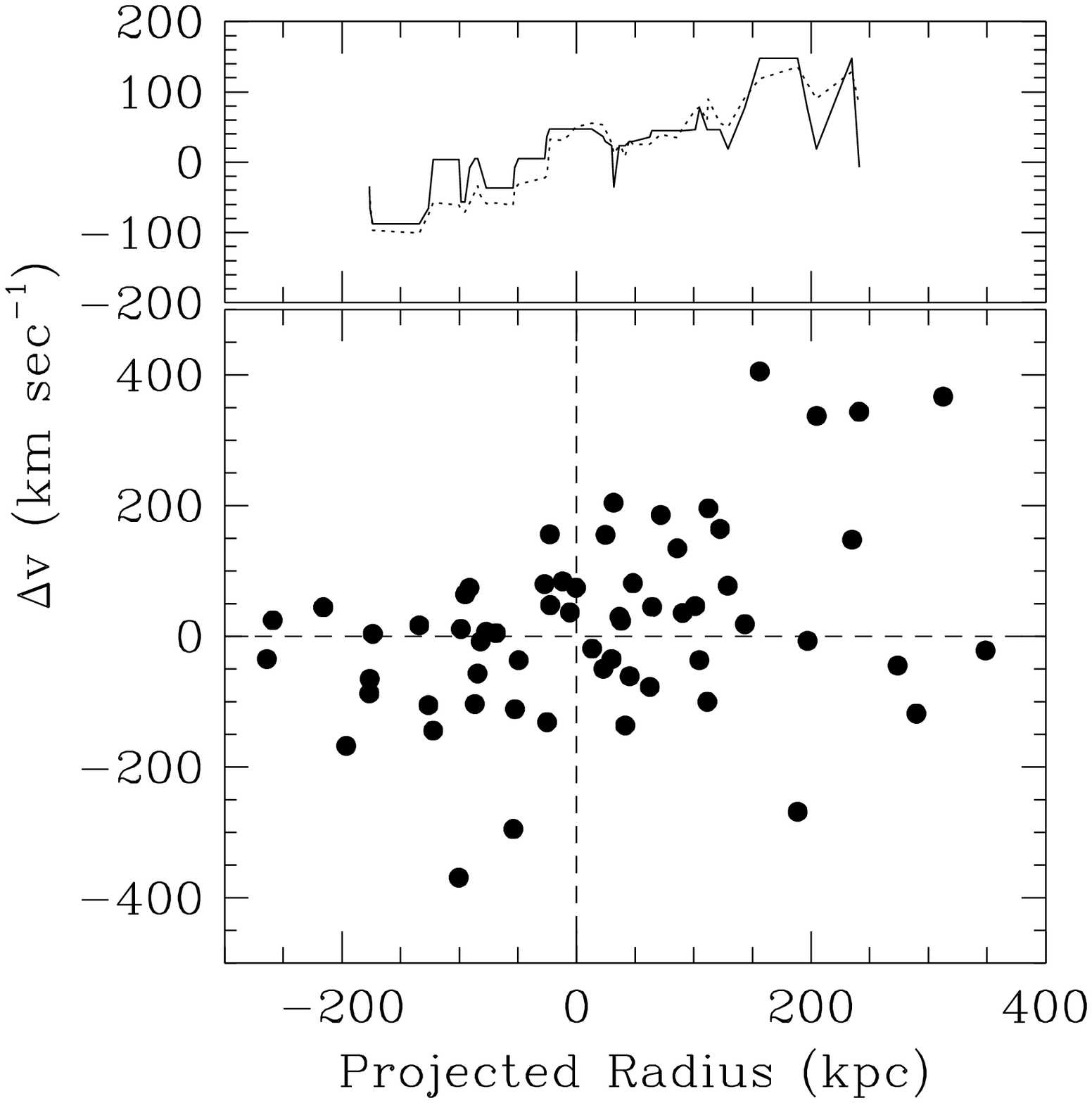}
\end{figure}
\end{document}